\documentclass[twocolumn,english,prl,amssymb,aps,superscriptaddress,showpacs,onecolumn,amsmath,showkeys,floatfix]{revtex4-1}
\usepackage[T1]{fontenc}
\usepackage[latin9]{inputenc}
\usepackage{geometry}
\usepackage[active]{srcltx}
\usepackage{textcomp}
\usepackage{amsmath}
\usepackage{graphicx}
\usepackage{color}
\usepackage{esint}
\usepackage{svg}
\makeatletter
\usepackage{babel}

\makeatother

\begin{document}
	
	\title {New method for residual amplitude modulation control in fibered optical experiments}
	\author{Maxime Descampeaux}
	\affiliation{Universit\'e Paris-Saclay, CNRS, ENS Paris-Saclay, CentraleSup\'elec, LuMIn, Gif-sur-Yvette, France}
	\affiliation {Thales Research and Technology, Palaiseau, France}
	\affiliation {Thales Avionics, Châtellerault, France}
	\author{Gilles Feugnet}
	\affiliation {Thales Research and Technology, Palaiseau, France}
	\author{Fabien Bretenaker}
	\affiliation{Universit\'e Paris-Saclay, CNRS, ENS Paris-Saclay, CentraleSup\'elec, LuMIn, Gif-sur-Yvette, France}

	\begin{abstract} 
When locking the frequency of a laser to an optical cavity resonance, the residual amplitude modulation (RAM), which accompanies the phase modulation necessary to build the error signal, is a major limitation to the frequency stability. We show that the popular method demonstrated by Wong and Hall to cancel this effect, based on the \textcolor{black}{measurement} of the RAM using an auxiliary detector, \textcolor{black}{ is limited in the case of optical setups exhibiting polarization dependent losses and an imperfect polarizer at the modulator output, such as guided-wave optical systems.}We propose and demonstrate a new method, using a single photodetector, to generate the two error signals and demonstrate its usefulness in the case of fibered systems.
	\end{abstract}
	
	
	\maketitle
	
\section{Introduction}

Laser frequency stabilization is mandatory for many applications, including optical gyroscopes \cite{Lei:20,Ravaille:19}, gravitational wave detection \cite{LIGO:2011}, spectroscopy \cite{Duong:18}, refractometry \cite{Egan:11}, etc. The tracking of the resonance frequency of a cavity can be achieved by observing the light transmitted or reflected by the cavity. An error signal is necessary for locking the system. It is very often obtained by a modulation-demodulation technique, like for example in the well-known Pound-Drever-Hall method \cite{Drever:83,Black:01}. One popular way to achieve such a frequency modulation consists in using an electro-optic phase modulator (PM) based on a lithium niobate (LiNbO$_3$) crystal, fed with an RF sinusoidal voltage. However, this phase modulation is very often accompanied by a spurious residual amplitude modulation (RAM) at the modulation frequency \cite{Zhang:14,Li:20}. In the demodulation process, this RAM cannot be separated from the useful signal induced by the phase modulation, and thus introduces a bias in the error signal, creating a detrimental frequency detuning from resonance once the servo-loop is closed. Moreover, this frequency offset varies in time, because it is related to two  phenomena that depend on the temperature: 1) a misorientation of the incident light polarization with respect to the principal axes of the birefringent LiNbO$_3$ crystal  and 2) the spurious etalon effects along the light path after the phase modulator \cite{Ishibashi:02,Shen:15,Whittaker:85}.

 The well established method to eliminate the RAM and its associated bias was introduced by Wong and Hall \cite{Wong:85}. It consists in splitting the beam into two beams after \textcolor{black}{a PM composed of a birefringent crystal sandwiched between two polarizers}. One of these beams is detected to measure the intensity modulation. After demodulation, one obtains an error signal set to zero by a servo-loop, which generates a DC voltage applied to the PM entrance to counteract. 
 This control loop locks the RAM amplitude to zero and compensates the natural birefringence of the PM crystal. The other part of the beam is then free from any RAM and can be used to lock the laser frequency on the reference cavity resonance. This method has been shown to be very efficient when there is no polarization dependent component on the two beam paths, for instance when both beams propagate in free space \cite{Duong:18,Zhang:14,Shen:15}. Several studies, based on this technique, propose complementary approaches to actively or passively compensate the RAM effect \cite{Bi:19,Shi:18,Jin:21,Li:16,Tai:16,Li:12}.

However, many experiments involving the locking of the laser frequency on a resonator use optical fibers and fibered components. This is the case for example for the passive resonant fiber-optic gyroscope \cite{Ravaille:19,Li:20}. In such a device, many fibered components, such as polarization maintaining fibers, fiber couplers, circulators, etc, exhibit polarization dependent losses and birefringence.  Furthermore, the polarization behavior of many of these components can be affected by mechanical stress, distortion, bending, inhomogeneity of the fiber, or temperature fluctuations \cite{Yao:03}. In this context, some authors have studied the RAM effect in the case of fiber coupled PMs \cite{Yu:16,Diehl:17,Li:14}, and  several techniques have been proposed to mitigate the RAM problem for gyroscope applications \cite{Jiang:19,Li:20,Liu:19}. However, one may wonder whether the well established and very efficient technique of Wong and Hall can be adapted to still operate in this context. The aim of the present paper is thus 1) to analyze the influence of the birefringence of the fibered components on the evolution of the RAM along the light path, and in particular between the detector that monitors the RAM and the resonator on which the laser frequency is locked and 2) to develop a new method for RAM control that makes use of the detector that provides the frequency locking signal to monitor the effect of the RAM on the frequency locking itself. 

To meet these goals, the paper is organized as follows. Section 2 explores, both theoretically and experimentally, the limits of the classical method introduced by Wong and Hall to cancel the RAM in the case of a fibered system\textcolor{black}{, taking into account the limited extinction ratio of the output polarizer}. Section 3 is devoted to the theoretical prediction of the fact that the same detector can be used to obtain an error signal for the laser frequency and an error signal for the RAM. To this aim, we derive an  analytical expression of an error signal including the effect of the RAM along the lines of the formalism introduced by Shen et al. \cite{Shen:15}. We present a new method to simultaneously perform laser frequency locking on the cavity resonance and RAM compensation with two separate loop filters exploiting two error signals derived from the same photodetector signal. We then compare these predictions with experiments and evaluate the performances of the usual and the new method  in Section 4.

\section{Investigation of the RAM effect with fibered components}

\subsection{Modeling}

\begin{figure}[h!]
\centering\includegraphics[width=7cm]{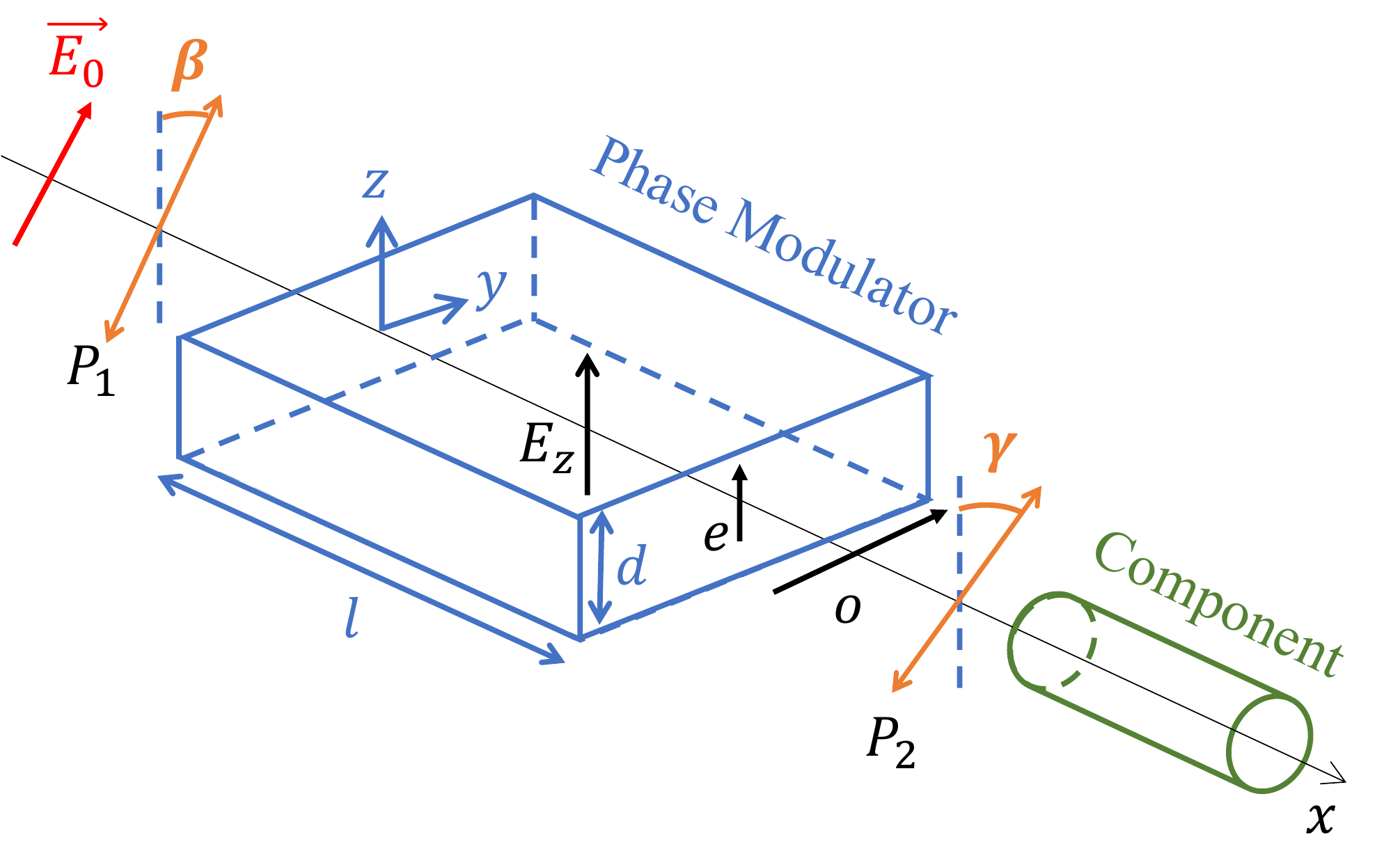}
\caption{Light travelling through a phase modulator (PM) located between two misaligned polarizers $P_1$ and $P_2$, followed by an extra fibered component. The o and e indices correspond respectively to the ordinary and extraordinary axes of the PM crystal of length $l$ and thickness $d$. The angles $\beta$ and $\gamma$ denote the orientation of the polarizers.}
\label{figure-1}
\end{figure}

As mentioned above, the Wong and Hall method to suppress the RAM requires two separate beams: one to lock the laser frequency on the reference and a second one to mitigate the RAM via a DC voltage applied to the phase modulator. In this section, we demonstrate that the RAM of the two beams can be different when one uses birefringent components with polarization dependant losses (PDL) and imperfect polarizers.

Figure\,\ref{figure-1} schematizes light at wavelength $\lambda$ propagating through a phase modulator sandwiched between two polarizers $P_1$ and $P_2$, followed by an extra fibered component. We use the Jones formalism \cite{Jones:41} to describe the propagation of the incident light field of amplitude $E_0$ at optical frequency $\omega_{\mathrm{L}}$ in this setup. This field  is initially fully linearly polarized after its passage through polarizer $P_1$. Its polarization is tilted by an angle $\beta$ with respect to the $z$ axis of the birefringent crystal along which the RF electric field $E_z$ is applied. This field creates a voltage $V_{\mathrm{RF}}\sin(\omega_{\mathrm{mod}}t)$ oscillating at the modulation frequency $\omega _{\mathrm{mod}}$ and associated with a DC offset voltage $V_{\mathrm{dc}}$. The light electric field after the modulator crystal is then given by:
\begin{equation}
E_{\mathrm{PM}}=E_{0}e^{i\omega_{\mathrm{L}} t}\begin{pmatrix} e^{i\Psi _{\mathrm{e}}}\cos\beta  \\e^{i\Psi _{\mathrm{o}}}\sin\beta  \end{pmatrix}\ ,
\label{eq-E-PM}
\end{equation}
where $\omega_{\mathrm{L}}$ is the laser frequency. The phase shifts $\Psi _{\mathrm{o}}$ and $\Psi _{\mathrm{e}}$ experienced by the components of the light field oriented along the ordinary and extraordinary crystal directions are given by

\begin{align}
\Psi _{\mathrm{o,e}} & = - \frac{\pi l}{\lambda d}r_{\mathrm{o,e}}n_{\mathrm{o,e}}^{3} V_{\mathrm{RF}}\sin(\omega _{\mathrm{mod}}t)-\frac{\pi l}{\lambda d}r_{\mathrm{o,e}}n_{\mathrm{o,e}}^{3} V_{\mathrm{dc}} + \frac{2\pi l}{\lambda }n_{\mathrm{o,e}}\nonumber  \\ & = \delta _{\mathrm{o,e}}\sin(\omega _{\mathrm{mod}}t)+ \phi _{\mathrm{o,e}}^{\mathrm{dc}} + \phi _{\mathrm{o,e}}^{\mathrm{n}}\ .\label{eq-total-phase-shift}
\end{align}
In this expression, the ordinary and extraordinary \textcolor{black}{refractive indexes $n_{\mathrm{o,e}}$, the electro-optic coefficients $r_{\mathrm{o,e}}$}, the modulation amplitudes $\delta _{\mathrm{o,e}}$, the DC electro-optically induced phase shifts $\phi _{\mathrm{o,e}}^{\mathrm{dc}}$ and the natural phase shifts $\phi _{\mathrm{o,e}}^{\mathrm{n}}$ depend on the temperature, the mechanical stress and the laser frequency fluctuations.

We then suppose that the second polarizer $P_2$ has an amplitude extinction ratio $\rho$ and is rotated by an angle $\gamma$ with respect to the $z$ axis of the crystal. After this polarizer, the beam is supposed to propagate through several components (fibers, couplers, circulators, polarizers...) that can be described by their Jones matrices that take into account their polarization dependent losses (PDL) and their birefringences \cite{Jones:41,Hurwitz:41,Chen:82,Tentori:13}.  Some of those matrices, for instance for a coupler, are complicated and would lead to calculations that would be outside the scope of this paper. In order to give a qualitative understanding of the considered phenomena, we consider the simple case of an optical fiber, as can be seen at the exit of polarizer $P_2$ in Fig.\,\ref{figure-1}. The electric field $E_{\mathrm{T}}$ at the output of the entire setup of Fig.\,\ref{figure-1} is then given by:

\begin{equation}
E_{\mathrm{T}} =   Q \times P \times R \times E_{\mathrm{PM}}\ , \label{eq-E-T-lineaire}
\end{equation}

with

\begin{equation}
\begin{aligned}
P=\begin{pmatrix} 1-\rho & 0 \\0 & \rho \end{pmatrix}, \quad
R=\begin{pmatrix} \cos\gamma & \sin\gamma \\-\sin\gamma & \cos\gamma \end{pmatrix}, \quad
Q=\begin{pmatrix} u_{11} & u_{12} \\u_{21} & u_{22} \end{pmatrix}\ .
\end{aligned}
\label{eq-P-R-Q}
\end{equation}

The Jones matrices $P$, $R$, and $Q$ hold for polarizer $P_2$, the rotation of angle $\gamma$, and the fibered component following $P_2$, respectively. The coefficients $u_{i,j}$ are in general complex. This leads to the following expression of $E_{\mathrm{T}}$:

\begin{equation}
\begin{aligned}
E_{\mathrm{T}} & = E_{0}e^{i\omega_0 t}\cdot \begin{pmatrix} 
[u_{11}(1-\rho)b-u_{12}\rho \hat{b}]e^{i\Psi _{e}} + [u_{11}(1-\rho)a+u_{12}\rho \hat{a}]e^{i\Psi _{o}}\\
[u_{21}(1-\rho)b-u_{22}\rho \hat{b}]e^{i\Psi _{e}} + [u_{21}(1-\rho)a+u_{22}\rho \hat{a}]e^{i\Psi _{o}}
\end{pmatrix}\ ,
\end{aligned}
\label{eq-E-T}
\end{equation}

\noindent where $a=\sin\beta\,\sin\gamma$, $b=\cos\beta\,\cos\gamma$, $\hat{a}=\sin\beta\,\cos\gamma$, and $\hat{b}=\cos\beta\,\sin\gamma$ are related to the misalignments between the modulator crystal and the two polarizers $P_1$ and $P_2$. \\

\subsubsection{Case of propagation in free space \textcolor{black}{after a perfect output polarizer}}
The case of propagation in free space after $P_2$ ($u_{11}=u_{22}=1$, $u_{12}=u_{21}=0$) and supposing polarizer $P_2$ to be perfect ($\rho =0$) leads to the well known results of Wong and Hall in Ref. \cite{Wong:85}. Indeed, in this case the transmitted electric field is given by 
\begin{equation}
E_{\mathrm{T}} =E_{0}\,e^{i\omega_{\mathrm{L}} t} [ be^{i\Psi _{\mathrm{e}}} + ae^{i\Psi _{\mathrm{o}}}  ]\ ,\label{eq-Wong-and-Hall-1}
\end{equation}
from which one can extract the component of the intensity $I_{\mathrm{T}}=|E_{\mathrm{T}}|^2$ modulated at frequency $\omega_{\mathrm{mod}}$:
\begin{equation}
I_{\mathrm{T}}(\omega_{\mathrm{mod}})=-4\,ab\left |E_{0}\right |^{2} J_{1}(M)\sin(\omega_{\mathrm{mod}} t)\sin(\Delta \phi)\ .
\label{eq-Wong-and-Hall-2}
\end{equation}
The quantity $M=\delta _{\mathrm{e}} -\delta _{\mathrm{o}}$ is the modulation amplitude, $J_1$ is the first order Bessel function, and $\Delta \phi = \Delta\phi ^{\mathrm{dc}} +\Delta\phi ^{\mathrm{n}}= (\phi _{\mathrm{e}}^{\mathrm{dc}} - \phi _{\mathrm{o}}^{\mathrm{dc}}) + (\phi _{\mathrm{e}}^{\mathrm{n}} - \phi _{\mathrm{o}}^{\mathrm{n}})$ is the crystal retardance. The technique developed by Wong and Hall consists in achieving $\sin(\Delta \phi)=0$ by adjusting the DC voltage $V_{\mathrm{dc}}$ applied to the crystal in such a way that its natural birefringence $\Delta\phi ^{\mathrm{n}}$ is compensated by $\Delta\phi ^{\mathrm{dc}}$. As shown by Eq.\,(\ref{eq-Wong-and-Hall-2}), this permits to cancel the intensity modulation at $\omega _{\mathrm{mod}}$.


\subsubsection{\textcolor{black}{Case of an imperfect polarizer followed by elements without polarization dependent losses}}
\textcolor{black}{This result can be easily generalized to the case of propagation in birefringent components without polarization dependent losses (PDL) after $P_2$, for which matrix $Q$ can, to a factor, be taken as special unitary \cite{Jones:41,Hurwitz:41}:}
\begin{equation}
Q_{\mathrm{lossless}}=    \begin{pmatrix} u_{11} & u_{12} \\ - u_{12}^* & u_{11}^* \end{pmatrix}\ ,
\end{equation}
with $\det(Q_{\mathrm{lossless}})=|u_{11}|^2+|u_{12}|^2=1$. 
Starting from Eq.\,(\ref{eq-E-T}), the part of the output intensity modulated at frequency $\omega_{\mathrm{mod}}$ becomes:
\begin{equation}
I_{\mathrm{T}}(\omega _{\mathrm{mod}})=-4 \left [ (ab(1-\rho)^2 - \hat{a}\hat{b}\rho^2 \right ]\left | E_{0} \right |^{2} J_{1}(M)\sin(\omega _{\mathrm{mod}} t)\sin(\Delta \phi)\ .\label{eq-no-PDL}
\end{equation}
The important point here is that this expression contains the same factor $\sin(\Delta \phi)$ as Eq.\,(\ref{eq-Wong-and-Hall-2}), showing that the method developed by Wong and Hall still works to cancel the RAM \textcolor{black}{even with an imperfect polarizer $P_2$ ($\rho \neq 0$)}.
%
%

\subsubsection{\textcolor{black}{Case of an imperfect polarizer followed by elements with polarization dependent losses}} \label{Sec2.1.3}
We finally consider the case where light propagates through components exhibiting polarization dependant losses after polarizer $P_2$. This would for instance be the case of a fiber coupler whose splitting ratio depends on the polarization. For simplicity, we focus on the simple case where the PDL consist in a perfect polarizer following some loss-free birefringent components describe by some unitary matrix  $Q_{\mathrm{lossless}}$. In this simple case the matrix $Q$ of Eq.(\ref{eq-E-T-lineaire}) is noted $Q_{\mathrm{PDL}}$ and becomes:
\begin{equation}
Q_{\mathrm{PDL}} = P_{\rho=0} \cdot Q_{\mathrm{lossless}} 
= \begin{pmatrix} 1 & 0 \\0 & 0 \end{pmatrix}
\cdot \begin{pmatrix} u_{11} & u_{12} \\ - u_{12}^* & u_{11}^* \end{pmatrix} = \begin{pmatrix} u_{11} & u_{12} \\ 0 & 0 \end{pmatrix}\ ,\label{eq-Q-PDL}
\end{equation}
Then electric field $E_{\mathrm{T}}$ and the modulated part $I_{\mathrm{T}}(\omega _{\mathrm{mod}})$ of the intensity become
\begin{equation}
\begin{aligned}
E_{\mathrm{T}} & =E_{0}e^{i\omega_{\mathrm{L}} t}\Big [ |b'|e^{i\phi _{b'}}e^{i\Psi _{\mathrm{e}}} + |a'|e^{i\phi _{a'}}e^{i\Psi _{\mathrm{o}}} \Big ]\ ,
\end{aligned}
\label{eq-E-PDL}
\end{equation}
\begin{equation}
I_{\mathrm{T}}(\omega _{\mathrm{mod}})=-4 |a'||b'| \left | E_{0} \right |^{2} J_{1}(M)\sin(\omega _{\mathrm{mod}} t)\sin(\Delta \phi + \phi _{b'} - \phi_{a'} )\ ,\label{eq-I-T-PDL}
\end{equation}
with
\begin{equation}
\left\{\begin{matrix}
a'=[u_{11}(1-\rho)a+u_{12}\rho \hat{a}]=|a'|e^{i\phi _{a'}}\ ,\\ 
b'=[u_{11}(1-\rho)b-u_{12}\rho \hat{b}]=|b'|e^{i\phi _{b'}}\ .
\end{matrix}\right.
\label{eq-B-A}
\end{equation}

One can notice that the condition for the RAM to vanish now becomes $\sin(\Delta \phi + \phi _{b'} - \phi_{a'} )=0$. As soon as the polarizer $P_{2}$ is not perfect, i. e. $\rho\neq0$, it thus depends on $\phi _{b'}$ and $\phi_{a'}$, i. e. on the propagation in the components \textcolor{black}{exhibiting PDL} that follow $P_2$. 
This simple model thus shows that RAM can occur after any component presenting polarization dependent losses following the phase modulator, as it was already mentioned in Ref. \cite{Gruning:19}. The results of the RAM measurement can then depend on the location along the optical circuit at which this measurement is performed. In particular, this means that, in the classical Wong and Hall technique to eliminate the RAM, suppressing the RAM at the position of one photodetector does not necessarily mean that the RAM is cancelled on the detector used to lock the laser frequency. Consequently, \textcolor{black}{with an imperfect polarizer $P_2$, in systems exhibiting PDL}, the Wong and Hall technique does not guarantee that spurious RAM induced frequency biases might not occur. 

Before proposing a method that will not face this issue, we check these predictions in the following subsection.


\subsection{Observation of RAM after propagation in a fibered component}
In the preceding subsection, we have predicted that propagation in fibered components exhibiting polarization dependent losses can change the RAM associated with the phase modulation. The present section reports an experiment aiming at testing this prediction, in which we compare two RAM measurements performed at two different points of a fibered optical circuit. The experimental setup designed to meet this goal is schematized in Fig.\,\ref{figure-2}. The details of the instruments used in this setup are given in the Supplemental. The laser light at $\lambda=1550\,\mathrm{nm}$ is phase modulated at $\omega_{\mathrm{mod}}/2\pi=70\,\mathrm{kHz}$ (the choice of this frequency is discussed in Section \ref{section4.1}) by the integrated optics LiNbO$_3$ PM and propagates in a polarization maintaining fiber (PMF) followed by a  polarisation maintaining 50:50 coupler C. This coupler, whose splitting ratio is polarization dependent, plays the role of the polarization dependent losses of the preceding subsection. The incident beam is thus split into two beams whose powers are monitored by two photodetectors PD$_1$ and PD$_2$. The components of the two signals at $\omega_{\mathrm{mod}}$ are then demodulated by the mixers M$_1$ and M$_2$ with the same phase by using a digital lock-in amplifier from Zurich Instruments. This leads to the two signals $V_{\mathrm{RAM1}}$ and $V_{\mathrm{RAM2}}$. The switch S then allows to select which demodulated signal is used as an error signal to cancel the RAM.  The loop filter LF contains a first-order 1~kHz bandwidth low-pass filter followed by a PID controller. The filtered loop signal is applied to the DC entrance of the PM.


\begin{figure}[h!]
\centering\includegraphics[width=7cm]{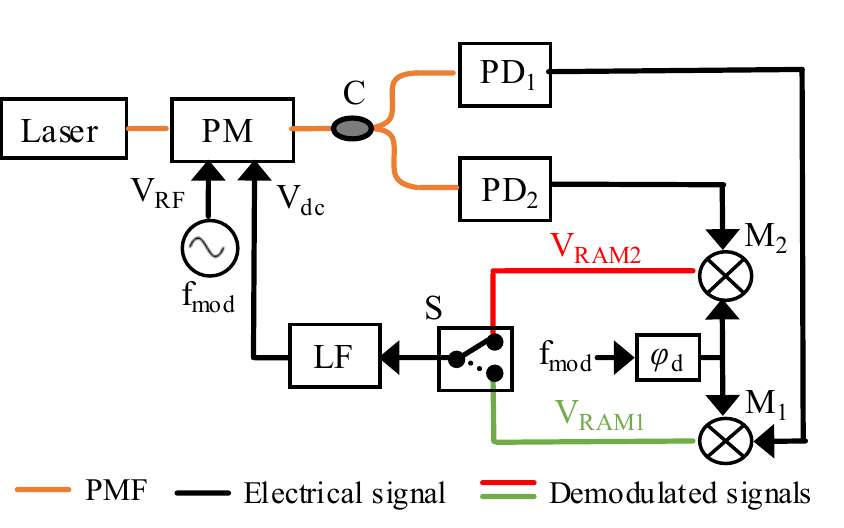}
\caption{Schematic diagram of the RAM monitoring  with two different photodetectors. C: coupler, PD: photodetector, M: mixer, S: switch, LF: loop filter.}
\label{figure-2}
\end{figure}

\begin{figure}[h!]
\centering\includegraphics[width=10cm]{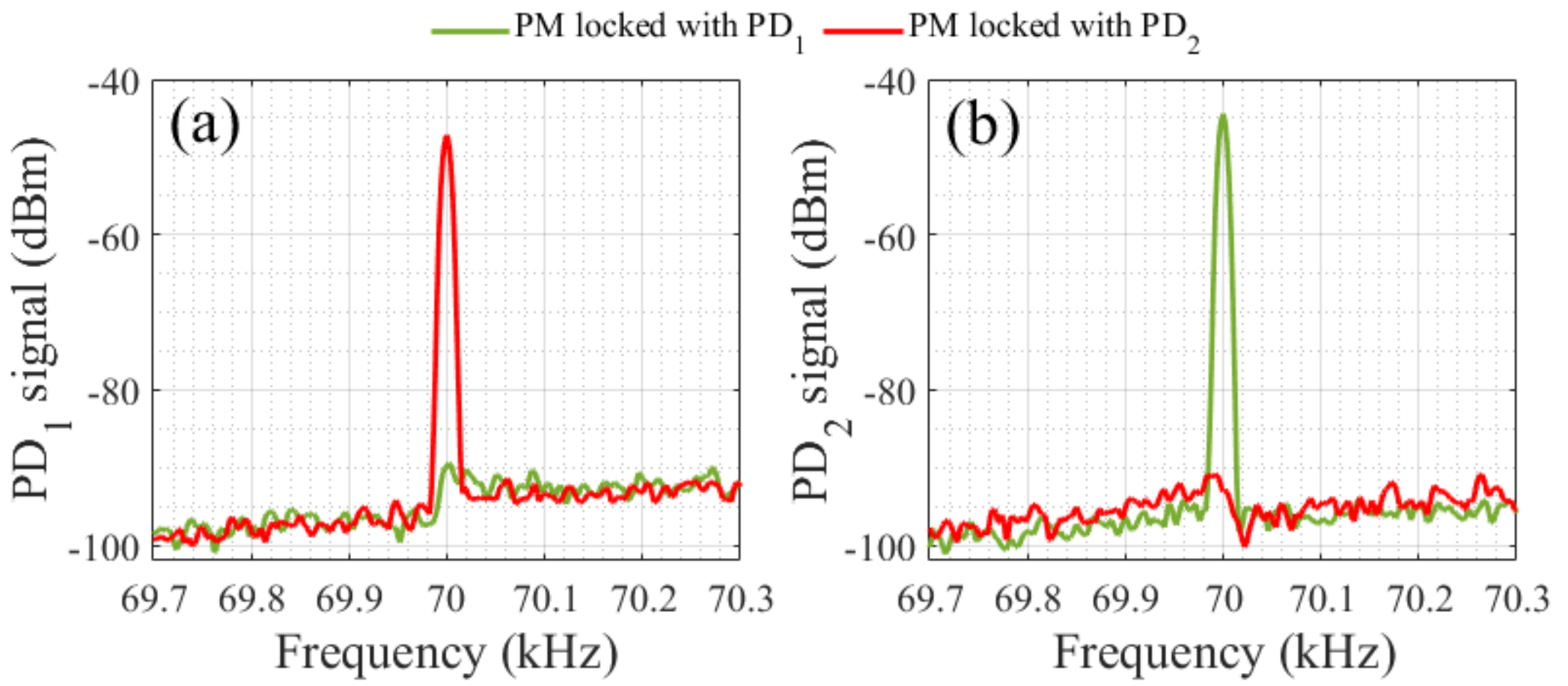}
\caption{Spectra of the photodetector signals of (a) PD$_1$ and (b) PD$_2$ while locking the phase modulator (PM) with the demodulated signal $\mathrm{V_{RAM1}}$ from PD$_1$ or $\mathrm{V_{RAM2}}$ from PD$_2$ at the same arbitrary demodulation phase $\varphi_{\mathrm{d}}$. RBW=10Hz.}
\label{figure-3}
\end{figure}

Figure \ref{figure-3} shows the spectra of the photodiode signals produced by (a) PD$_1$ and (b) PD$_2$ around $f_{\mathrm{mod}}=70\,\mathrm{kHz}$. The green lines (resp. red lines) are recorded when the signal $\mathrm{V_{RAM1}}$ from PD$_1$ (resp. $\mathrm{V_{RAM2}}$ from PD$_2$) is used as an error signal to reduce the RAM. 
Finally, two different PMs are used: a first one followed by a polarizer  at its output \textcolor{black}{(as in the original Wong and Hall experiments)} and a second one that does not contain an output polarizer. The results of Fig.\,\ref{figure-3} and all the other results of this paper correspond to the latter one. \textcolor{black}{We choose to present the results of the experiments using the PM without any output polarizer because the RAM is much larger in this case (while the results for a PM followed by a polarizer are presented in the Supplemental). This case is thus a better illustration of the efficiency of the RAM control method proposed here, which works in both cases.}

One can clearly see that the RAM peak at $f_{\mathrm{mod}}=70\,\mathrm{kHz}$ is perfectly canceled when it is measured with the photodiode whose signal is used for the servo-loop, while it is still present in the other photodiode signal. For example, by comparing the green signals in Figs.\,\ref{figure-3}(a) and \ref{figure-3}(b), it is clear that the RAM is perfectly canceled at the position of PD$_1$ while it is still clearly present at the position of PD$_2$.


This simple experiment thus perfectly confirms the analysis of the preceding subsection and evidences the limitations of the classical method \textcolor{black}{based on two photodiodes, in the case of an imperfect polarizer located after the PM and followed by fibered components}. It is thus necessary to find a way to extract the error signal for the frequency locking of the laser and the error signal for the RAM cancellation from the same photodetector. This is the aim of the following section.


\section{Theoretical extraction of two error signals from a single detector}\label{sec3}

In this section, we theoretically investigate the possibility to extract two different error signals from a single photodetector, one to lock the laser frequency at resonance and one to cancel the RAM. To this aim, the first subsection is dedicated to a calculation of the response of any type of cavity resonance, either in transmission or reflection, to a phase and intensity modulated light signal. The second subsection aims at deriving the two error signals and predicting the possibility to use them to cancel the RAM and lock the frequency simultaneously.


\subsection{General expression of the demodulated signal}

This subsection is dedicated to the derivation of a general expression of the  signal at the output of a cavity, probed either in reflection or transmission.  The cavity free spectral range is noted $\Delta \nu _{\mathrm{FSR}}$. Based on Eqs. (\ref{eq-total-phase-shift}) and (\ref{eq-E-PDL}), the modulated field $E_{\mathrm{T}}$ incident on the cavity can be expanded according to its frequency components centered around the laser frequency $\omega_{\mathrm{L}}$. The different sidebands are separated by  integer multiples of the modulation frequency $\omega _{\mathrm{mod}}$. The field $E_{\mathrm{out}}$ at the output of the cavity is obtained by multiplying each frequency component at $\omega_{\mathrm{L}} +k \omega _{\mathrm{mod}}$ of $E_{\mathrm{T}}$ by the transfer function $F(\omega_{\mathrm{L}} +k \omega _{\mathrm{mod}})$ of  the cavity taken at the corresponding frequency, leading to the same expression as in Ref. \cite{Shen:15}:

\begin{equation}
E_{\mathrm{out}}  = E_{0}\sum_{k=-\infty }^{+\infty }\left[a'e^{i\phi _{\mathrm{o}}}J_k ^{\mathrm{o}}+ b'e^{i\phi _{\mathrm{e}}}J_k^{\mathrm{e}}\right] F(\omega_{\mathrm{L}} +k \omega _{\mathrm{mod}})  e^{i(\omega_{\mathrm{L}} +k \omega _{\mathrm{mod}})t}\ ,\label{eq-E-out}
\end{equation}
where $J_k ^{\mathrm{o,e}}=J_k(\delta _{\mathrm{o,e}})$ with $J_k$ the Bessel function of order $k$, $\phi _{\mathrm{o,e}}=\phi _{\mathrm{o,e}}^{\mathrm{dc}} + \phi _{\mathrm{o,e}}^{\mathrm{n}}$ (see Eq.\,\ref{eq-total-phase-shift}), and $a'$ and $b'$ are the complex numbers previously introduced in Eq.(\ref{eq-B-A}). The details of the transfer function of the cavity are given in the Supplemental. 
The output intensity  $I_{\mathrm{out}}=E_{\mathrm{out}}E_{\mathrm{out}}^*$ is then detected by a photodiode and demodulated at frequency $\omega_{\mathrm{mod}}$ with a demodulation phase $\varphi _{\mathrm{d}}$, leading to the following demodulated signal, whose complete derivation is given in the Supplemental:
\begin{equation}
V_{\mathrm{\varepsilon}}(\varphi _{\mathrm{d}}) = G I_0 \sum_{k=0}^{+\infty }   \left[\mathrm{Re}(b_{k})\sin(\varphi _{\mathrm{d}}) +\mathrm{Im}(b_{k})\cos(\varphi _{\mathrm{d}}) \right]\ , 
\label{eq-Verr-n}
\end{equation}
where $I_0$ is the input intensity, $G$ is the optoelectronic gain (in V/W) that contains the response of the detector, the gain of the amplifier, and the gain of the demodulation at frequency $\omega_{\mathrm{mod}}$. The coefficients in Eq.\,(\ref{eq-Verr-n}) are given by:
\begin{align}
b_{k} = & A_{k}F\left(\omega_{\mathrm{L}} +k \omega _{\mathrm{mod}}\right)F^*\left(\omega_{\mathrm{L}} +(k+1) \omega _{\mathrm{mod}}\right) \nonumber\\
& -A_{k}^*F\left(\omega_{\mathrm{L}} -(k+1) \omega _{\mathrm{mod}}\right)F^*\left(\omega_{\mathrm{L}} -k \omega _{\mathrm{mod}}\right)\ ,\label{eq-bk1-expression}
\end{align}
with
\begin{align}
A_{k}  = &  |a'|^2 J_k ^{\mathrm{o}} J_{k+1} ^{\mathrm{o}} + |b'|^2 J_k ^{\mathrm{e}} J_{k+1} ^{\mathrm{e}} + |a'b'| \left ( J_k ^{\mathrm{o}} J_{k+1} ^{\mathrm{e}} + J_k ^{\mathrm{e}} J_{k+1} ^{\mathrm{o}} \right ) \cos(\delta \phi)  \nonumber\\
&- i |a'b'| \left ( J_k ^{\mathrm{o}} J_{k+1} ^{\mathrm{e}} - J_k ^{\mathrm{e}} J_{k+1} ^{\mathrm{o}} \right ) \sin(\delta \phi)\ .\label{eq-Akp}
\end{align}
In this expression, we have introduced the quantity
\begin{equation}
\delta \phi=\Delta \phi + \phi _{b'} - \phi_{a'}\ ,\label{Eq18}
\end{equation}
which is the phase shift  that describes the amount of RAM introduced by the modulator in presence of birefringence and polarization dependent losses, as shown in Section \ref{Sec2.1.3}. 


As an illustration, we consider a ring cavity whose parameters correspond to the experiment that will be described later. It has a finesse of 90 and and a free spectral range (FSR)  $\Delta \nu _{\mathrm{FSR}}=13.6\,\mathrm{MHz}$, corresponding to a resonance  width $\Gamma\simeq150\,\mathrm{kHz}$.  The cavity is supposed to be probed in transmission, and the modulation frequency is chosen to be $f_{\mathrm{mod}}=\omega_{\mathrm{mod}}/2\pi=70\,\mathrm{kHz}$, i.~e. close to $\Gamma/2$ in order to optimize the slope of the error signal. For simplicity, we suppose here that $a'=a$ and $b'=b$. We also take  $\beta = \gamma = 20^{\circ}$. Such values are exaggerated with respect to real experimental situations, but aim at making the effect of the RAM more visible. Finally, we take the maximum value of $k$ to be 5 in the summation of Eq.\,(\ref{eq-Verr-n}).

\begin{figure}[h!]
\centering\includegraphics[width=13cm]{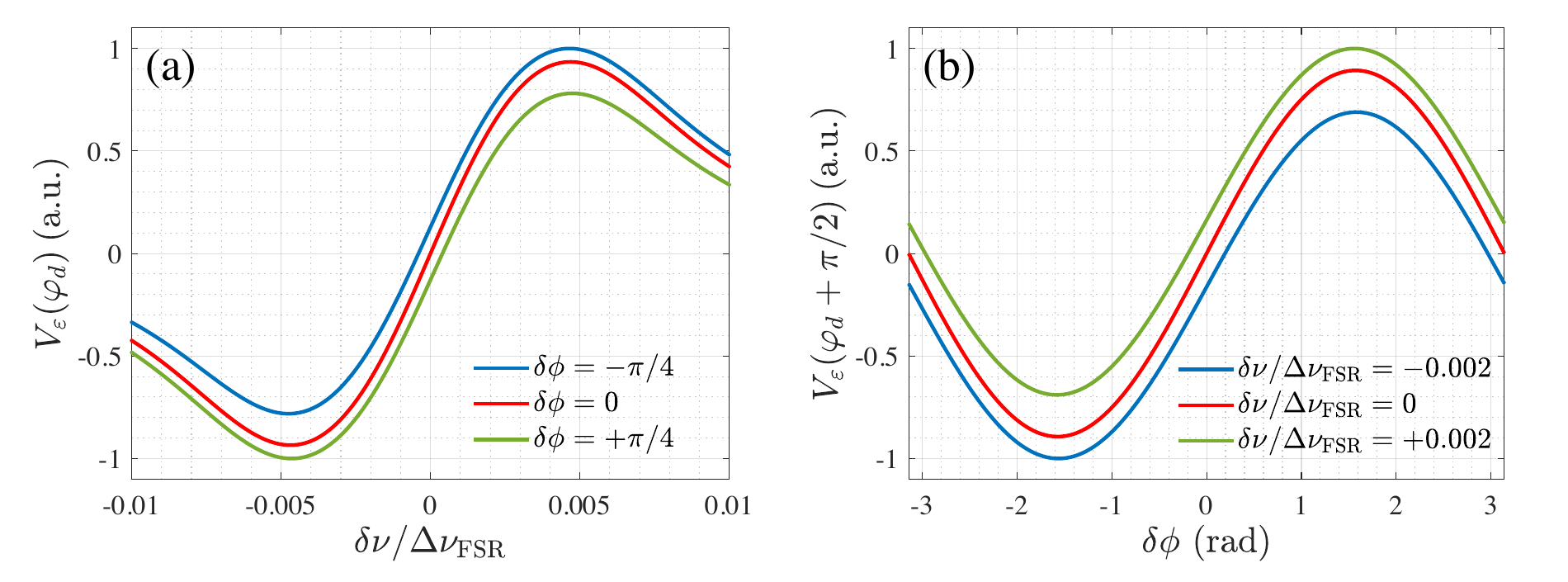}
\caption{(a) Computed evolution of the error signal $V_{\varepsilon}(\varphi_{\mathrm{d}})$ versus  detuning $\delta \nu$ of the laser from the cavity resonance frequency normalized to the FSR $\Delta \nu _{\mathrm{FSR}}$ for three values of $\delta \phi$.  (b) Evolution of  $V_{\varepsilon}(\varphi_{\mathrm{d}} + \pi /2)$ versus $\delta \phi$ for three values of $\delta \nu/\Delta \nu _{\mathrm{FSR}}$. The cavity is supposed to be probed in transmission with  $a'=a$, $b'=b$, $\beta = \gamma = 20^{\circ}$, $\delta _{\mathrm{o}}=0.3\,\mathrm{rad}$,  $\delta _{\mathrm{e}}=1\,\mathrm{rad}$, and $\varphi _{\mathrm{d}}=-1.555\,\mathrm{rad}$.}
\label{figure-4}
\end{figure}

We then optimize the value of the demodulation phase $\varphi _{\mathrm{d}}$ in order to maximize the evolution of the demodulated signal $V_{\varepsilon}$ versus  frequency detuning  $\delta\nu$, leading to $\varphi _{\mathrm{d}}=-1.555\,\mathrm{rad}$. 
The corresponding evolution of the demodulated signal $V_{\varepsilon}(\varphi_{\mathrm{d}})$ of Eq.\,(\ref{eq-Verr-n}) with  $\delta \nu / \Delta \nu _{\mathrm{FSR}}$  is reproduced in  Fig.\,\ref{figure-4}(a) for three different values of $\delta \phi$. As can be seen from this figure, for $\delta \phi=0$, the RAM and the associated frequency bias vanish: the demodulated signal is equal to zero when the field is exactly at resonance with the cavity, and evolves linearly around resonance. It thus constitutes a good error signal to lock the laser frequency. On the contrary, when $\delta \phi\neq 0$, the RAM creates a frequency offset. 

We now wonder if a second error signal can be obtained from the same photodiode voltage, that will allow us to lock $\delta \phi$ to 0, and thus cancel the RAM. To this aim, we plot the evolution of the  signal obtained by demodulating the photocurrent in quadrature, i. e., with the phase $\varphi _{\mathrm{d}}+\pi/2$, as a function of $\delta \phi$ (see Fig.\,\ref{figure-4}(b)). As can be seen from this figure, the new demodulated signal $V_{\varepsilon}(\varphi _{\mathrm{d}}+\pi/2)$ crosses 0 exactly for $\delta\phi=0$ when $\delta\nu=0$ and evolves linearly around this value. It thus constitutes a good error signal for the cancellation of the RAM. On the contrary, if a small bias $\delta\nu\neq0$ exists, then the  signal vanishes for $\delta\phi$ slightly different from 0.

The example explored in this subsection suggests that by demodulating the photocurrent with two different quadratures, one can obtain two error signals. The fact that these two error signals  simultaneously vanish  ensures that both $\delta\phi=0$ and $\delta\nu=0$ and that the laser frequency is locked on the cavity resonance without any RAM-induced offset. 

The next section generalizes this result by exploring theoretically how two signals demodulated with different phases permit to achieve the same result.



\subsection{Expressions of the linearized error signals}

An analytical expression of the two error signals useable to design the double servo-locking can be obtained by linearizing the demodulated signal of Eq.\,(\ref{eq-Verr-n}) for a laser detuning $\delta\nu\ll\Gamma$ and for a very small value of the RAM, i. e., $\delta\phi\ll 1$. The details of the linearization of the demodulated signal with respect to $\delta\omega=2\pi \delta\nu$ and $\delta\phi$ are given in the Supplemental and lead to:

%
%
%

\begin{equation}
 V_{\varepsilon} =  G I_0 \left ( \delta \phi \: D_{\delta \phi} -\delta \omega\:  D_{\delta \omega} \right )\ , 
\label{eq-Verr-linear}
\end{equation}
where the discriminator slopes are given by
\begin{align}
D_{\delta \phi} =    2 |a'b'| \sum_{k=0}^{+\infty }&\left( J_k^{\mathrm{o}} J_{k+1}^{\mathrm{e}}  -  J_k ^{\mathrm{e}} J_{k+1} ^{\mathrm{o}} \right )   \Big \{\mathrm{Im} \left[ F(\omega_{\mathrm{0}} +k \omega _{\mathrm{mod}})F(\omega_{\mathrm{0}} -(k+1) \omega _{\mathrm{mod}})\right] \sin\varphi_{\mathrm{d}}  \nonumber\\
&-\mathrm{Re} \left[ F(\omega_{\mathrm{0}} +k \omega _{\mathrm{mod}})F(\omega_{\mathrm{0}} -(k+1) \omega _{\mathrm{mod}}) \right] \cos\varphi_{\mathrm{d}}  \Big \}\ ,
\label{eq-slope-delta-phi}
\end{align}
and
\begin{align}
D_{\delta \omega} = & \ 2 \sum_{k=0}^{+\infty } \Big [ |a'|^2 J_k ^{\mathrm{o}} J_{k+1} ^{\mathrm{o}} + |b'|^2 J_k ^{\mathrm{e}} J_{k+1} ^{\mathrm{e}} + |a'b'| \left ( J_k ^{\mathrm{o}} J_{k+1} ^{\mathrm{e}} + J_k ^{\mathrm{e}} J_{k+1} ^{\mathrm{o}} \right ) \Big ] \nonumber \\ 
& \times \big \{ \mathrm{Re} \left [ F\left(\omega_{\mathrm{0}} +k \omega _{\mathrm{mod}}\right)   F'  (\omega _0-(k+1)\omega_{\mathrm{mod}}) - F(\omega_{\mathrm{0}} -(k+1) \omega _{\mathrm{mod}})   F'(\omega _0+k\omega_{\mathrm{mod}})\right ] \sin\varphi_{\mathrm{d}}   \nonumber
\\ 
& + \mathrm{Im} \left[ F(\omega_{\mathrm{0}} +k \omega _{\mathrm{\mathrm{mod}}})   F'(\omega _0-(k+1)\omega_{\mathrm{mod}}) - F(\omega_{\mathrm{0}} -(k+1) \omega _{\mathrm{mod}})  F'(\omega _0+k\omega_{\mathrm{mod}})\right ]\cos\varphi_{\mathrm{d}}  \big \}\ , 
\label{eq-slope-delta-nu}
\end{align}
with $F'=dF/d\omega$. 

Equation (\ref{eq-Verr-linear}) shows that in the presence of RAM ($\delta\phi\neq0$) the demodulated signal can be equal to zero even if $\delta\omega\neq0$ so if $\delta\nu\neq0$. In order to cancel both $\delta\phi$ and $\delta\nu$, one must thus demodulate the photodiode signal with two different phases $\varphi _{\mathrm{d1}}$ and $\varphi_{\mathrm{d2}}$, leading to two different error signals $V_{\varepsilon}(\varphi _{\mathrm{d1}})$ and $V_{\varepsilon}(\varphi _{\mathrm{d2}})$.
The cancellation of the frequency detuning $\delta\nu$ and of the RAM $\delta\phi$ can be obtained by locking these two error signals to 0, i.e., 
\begin{equation}\begin{aligned}
\left \{\begin{array}{rcl}
V_{\varepsilon}(\varphi _{\mathrm{d1}})= G I_0 \Big ( \delta \phi \: D_{\delta \phi}(\varphi _{\mathrm{d1}}) -\delta \omega\:  D_{\delta\omega}(\varphi _{\mathrm{d1}}) \Big ) =0\ ,
 \\V_{\varepsilon}(\varphi _{\mathrm{d2}})= G I_0 \Big ( \delta \phi \: D_{\delta \phi}(\varphi _{\mathrm{d2}}) -\delta \omega\:  D_{\delta \omega}(\varphi _{\mathrm{d2}}) \Big ) =0\ ,
\end{array} \right. 
\label{eq-two-Ve-zero}
\end{aligned}\end{equation}
provided the two equations are independent:
\begin{equation}
 D_{\delta \phi}(\varphi _{\mathrm{d1}})D_{\delta \omega}(\varphi _{\mathrm{d2}})-D_{\delta \phi}(\varphi _{\mathrm{d2}})D_{\delta \omega}(\varphi _{\mathrm{d1}})\neq 0\ .\label{Eq23}
\end{equation}
One condition sufficient to fulfil Eq.\,(\ref{Eq23}) is that  $\varphi_{\mathrm{d2}}=\varphi_{\mathrm{d1}}\pm\pi/2$, as can be seen by carefully analyzing Eqs.\,(\ref{eq-slope-delta-phi},\ref{eq-slope-delta-nu}).  \textcolor{black}{However this is not a necessary condition to cancel both the RAM and the frequency detuning simultaneously.}

\section{Experiment with RAM control method}\label{section4}
This section is devoted to the experimental implementation of the new servo-locking strategy predicted in the preceding section. To this aim, we first compare the predicted evolutions of the demodulated signals with experiments. We then use them to show that the RAM can indeed be cancelled by using the same photodiode signal as the one used to lock the frequency. Finally, in the last subsection, we show that the two servo-lockings can be made to operate simultaneously to lock the laser at cavity resonance while being immune to RAM-induced frequency offsets. 

\subsection{Observation of the two error signals}\label{section4.1}
\begin{figure}[h!]
\centering\includegraphics[width=8cm]{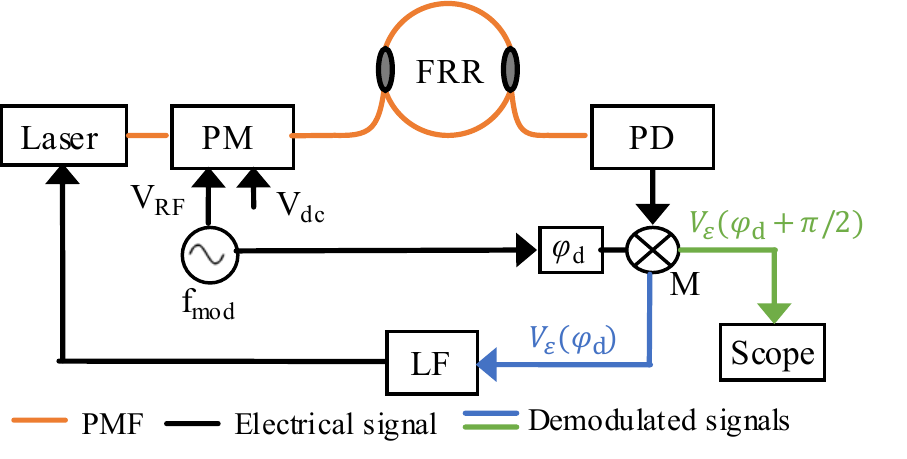}
\caption{Schematic diagram of the laser frequency locking on the FRR cavity. PD: photodetector, M: mixer, LF: loop filter.}
\label{figure-5}
\end{figure}
The experiment of Fig.\,\ref{figure-5} has been used to check the theoretical predictions. It uses a polarization maintaining single mode fiber ring resonator (FRR). This temperature stabilized $L=22\,\mathrm{m}$ long fiber ring cavity has a finesse of 90, a diameter of 8\,cm, a resonance width $\Gamma\simeq150\,\mathrm{kHz}$, and a free spectral range (FSR)  $\Delta \nu _{\mathrm{FSR}}=13.608\;\mathrm{MHz}$. The light from the laser is phase modulated at $f_{\mathrm{mod}}=70\,\mathrm{kHz}$. 

In a first experiment, we measure the error signals by applying a triangular voltage to the laser frequency modulation input and keeping the servo-loop open. The intensity transmitted by the resonator is demodulated at $f_{\mathrm{mod}}$ and the two output quadrature signals are filtered with a first-order low-pass filter with a 1~kHz bandwidth. 

\begin{figure}[h!]
\centering\includegraphics[width=9cm]{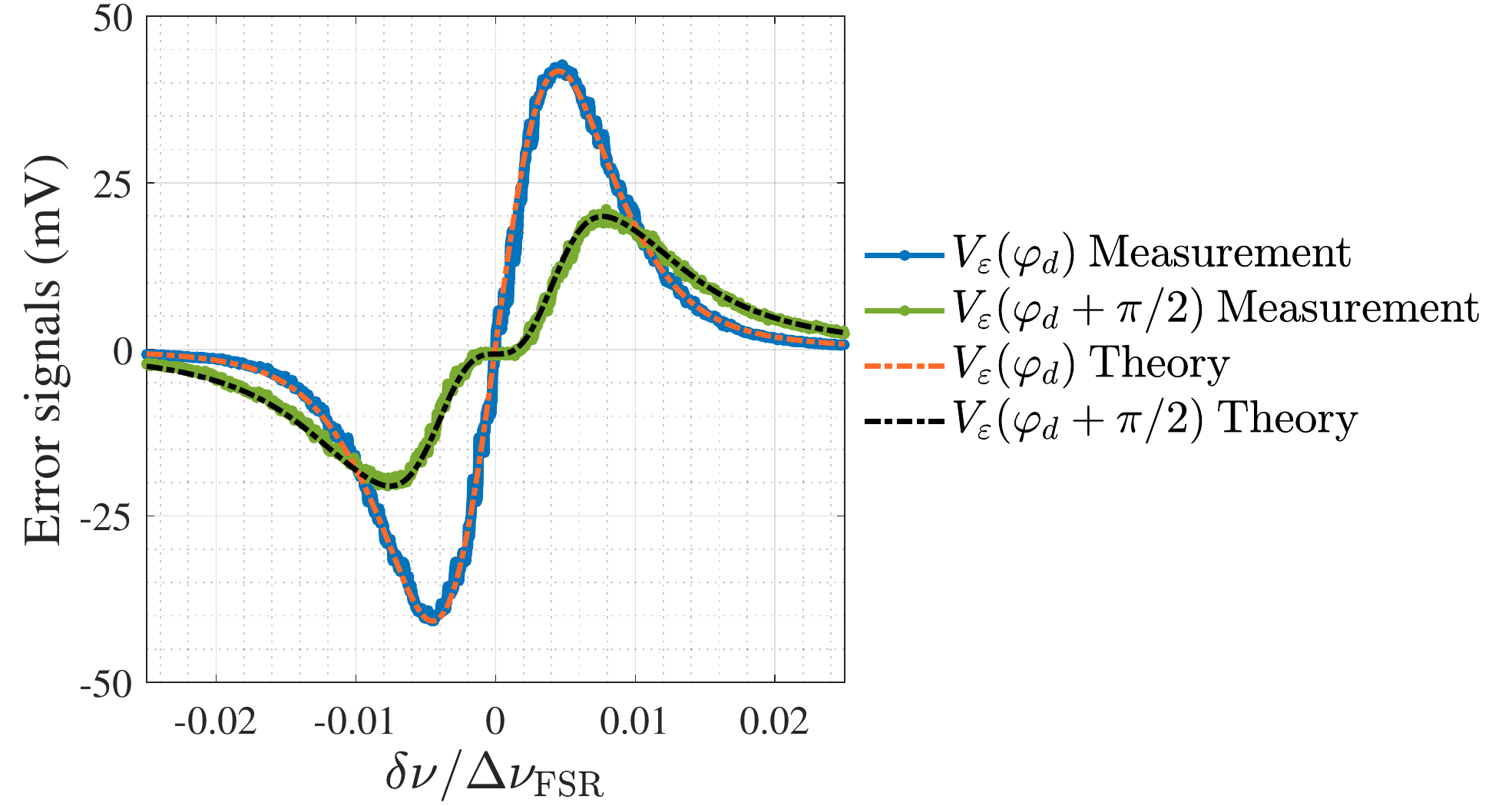}
\caption{Measured and fitted open loop  error signals $V_{\varepsilon}(\varphi_{\mathrm{d}})$ and $V_{\varepsilon}(\varphi_{\mathrm{d}}+\pi/2)$ against frequency detuning normalized to the FSR $\delta \nu / \Delta \nu _{\mathrm{FSR}}$  for $f_{\mathrm{mod}}=70$kHz with a demodulation phase $\varphi _{\mathrm{d}}$ giving a maximum frequency slope on $V_{\varepsilon}(\varphi _{\mathrm{d}})$.}
\label{figure-6}
\end{figure}

Figure\,\ref{figure-6} shows the evolution of these two demodulated signals as a function of the cavity detuning $\delta\nu$ normalized by the FSR. In this experiment the demodulation phase $\varphi_{\mathrm{d}}$ has been adjusted in order to maximize the slope of one of the demodulated signals \textcolor{black}{(the blue one in Fig.\,\ref{figure-6}) around $\delta\nu=0$. This signal provides a good discrimination to lock the frequency. On the contrary, we can see that the other quadrature exhibits a plateau around resonance and is thus not suitable to lock the frequency. But, as we will see later, it provides a good error signal for the RAM control}. Figure\,\ref{figure-6} compares the experimental signals demodulated at $\varphi_{\mathrm{d}}$ and $\varphi_{\mathrm{d}}+\pi/2$ with their theoretical counterparts, calculated from Eq.\,(\ref{eq-Verr-n}). To get these curves, we adjusted the parameter values $\beta = \gamma = 5^{\circ} $ and $\delta \phi = \pi/8$. An excellent agreement is obtained between theory and experiments, and the values of these parameters are coherent with the alignment tolerances of PMF connectors.


The results of Fig.\,\ref{figure-4} suggest that, with this choice of $\varphi_{\mathrm{d}}$, the demodulated signal $V_{\varepsilon}(\varphi_{\mathrm{d}})$ can be used as an error signal for the frequency \textcolor{black}{with a high discrimination slope}, while $V_{\varepsilon}(\varphi_{\mathrm{d}}+\pi/2)$ \textcolor{black}{could} be used to monitor and eventually cancel the RAM.
\begin{figure}[h!]
\centering\includegraphics[width=9cm]{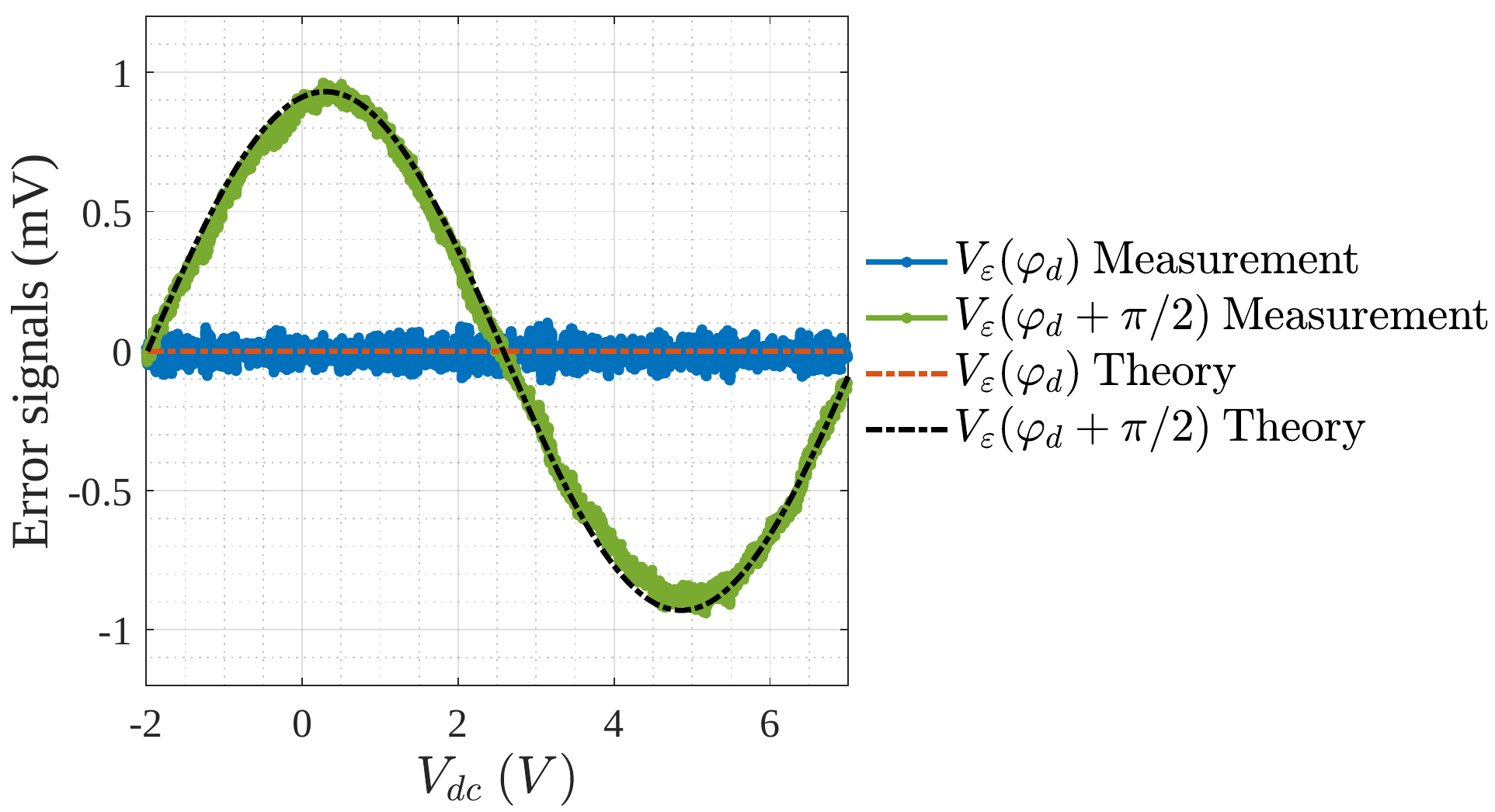}
\caption{Measured and fitted error signals $V_{\varepsilon}(\varphi_{\mathrm{d}})$ and $V_{\varepsilon}(\varphi_{\mathrm{d}}+\pi/2)$ against DC voltage $V_{\mathrm{dc}}$ applied to the PM for $f_{\mathrm{mod}}=70$kHz with a demodulation phase $\varphi _{\mathrm{d}}$ giving a maximum frequency slope on $V_{\varepsilon}(\varphi _{\mathrm{d}})$. Only the frequency locking servo-loop is closed.}
\label{figure-7}
\end{figure}
To test this possibility, we close the frequency locking servo-loop as shown in Fig.\,\ref{figure-5}, and we monitor the evolution of the two demodulated signals as a function of the DC voltage $V_{\mathrm{dc}}$ applied to the PM to adjust the RAM amplitude. The corresponding results are reproduced in Fig.\,\ref{figure-7}. These results show that the servo-loop maintains the error signal $V_{\varepsilon}(\varphi _{\mathrm{d}})$ at zero while the error signal $V_{\varepsilon}(\varphi _{\mathrm{d}}+\pi/2)$ evolves sinusoidally with $V_{\mathrm{dc}}$. This is consistent with the fact that 
$V_{\varepsilon}(\varphi _{\mathrm{d}}+\pi/2)$ monitors the RAM amplitude, which is proportional to $\sin(\textcolor{black}{\delta \phi})$ according to Eqs.\,(\ref{eq-Verr-n}-\ref{eq-Akp}). Again, the theoretical curves perfectly agree with the measured ones. This result shows that the two demodulated signals provide two error signals. In principle, according to Sec.\,\ref{sec3}, locking $V_{\varepsilon}(\varphi _{\mathrm{d}}+\pi/2)$ at zero by controlling $V_{\mathrm{dc}}$ should cancel the RAM at the detector position and thus ensure that $V_{\varepsilon}(\varphi _{\mathrm{d}})=0$ corresponds to $\delta\nu=0.$

\subsection{Observation of RAM cancellation}
\begin{figure}[h!]
\centering\includegraphics[width=7cm]{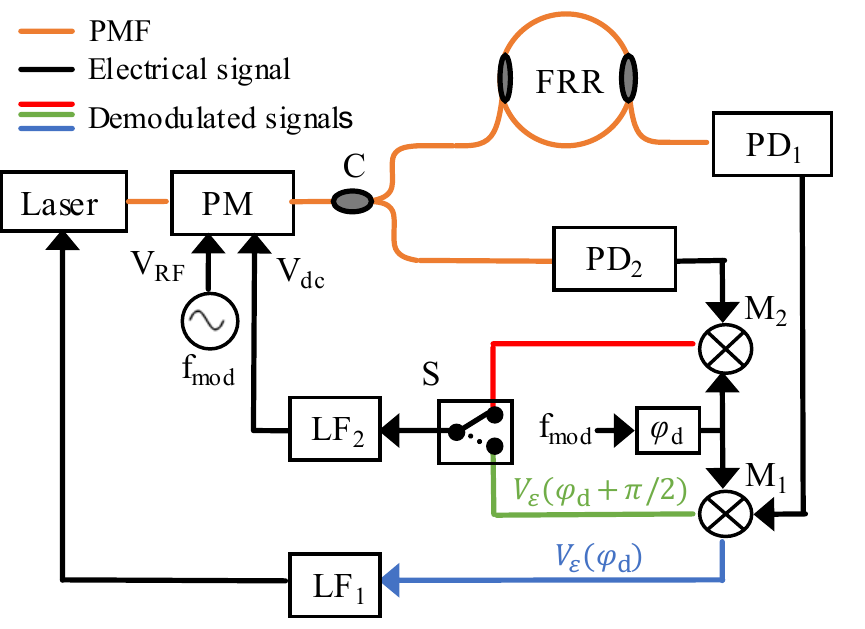}
\caption{Schematic diagram of RAM cancellation method by locking the PM either with $V_{\varepsilon}(\varphi _{\mathrm{d}}+\pi/2)$ from PD$_1$ (new method, green wire) or with the demodulated signal from PD$_2$ (classical method, red wire). The laser frequency is locked by using $V_{\varepsilon}(\varphi _{\mathrm{d}})$ from PD$_1$ in both cases (blue wire). C: coupler, PD: photodetector, M: mixer, S: switch, LF: loop filter.}
\label{figure-8}
\end{figure}

To confirm that $V_{\varepsilon}(\varphi _{\mathrm{d}}+\pi/2)=0$ indeed corresponds to a cancellation of the RAM on the detector,  we perform the experiment schematized in Fig.\,\ref{figure-8}. This setup uses the same resonator as in Fig.\,\ref{figure-5}. Detector PD$_1$ monitors the light transmitted by this resonator. An extra detector PD$_2$ monitors the light intensity before the resonator. In this experiment, the laser frequency is locked by demodulating with a phase $\varphi_{\mathrm{d}}$ the voltage provided by PD$_1$ and applying the resulting error signal to the laser. Besides, the RAM is monitored and/or canceled by demodulating the signal provided either by PD$_1$ (with a phase $\varphi_{\mathrm{d}} +\pi/2$) or PD$_2$ (with a phase $\varphi_{\mathrm{d}}$).
\begin{figure}[h!]
\centering\includegraphics[width=7cm]{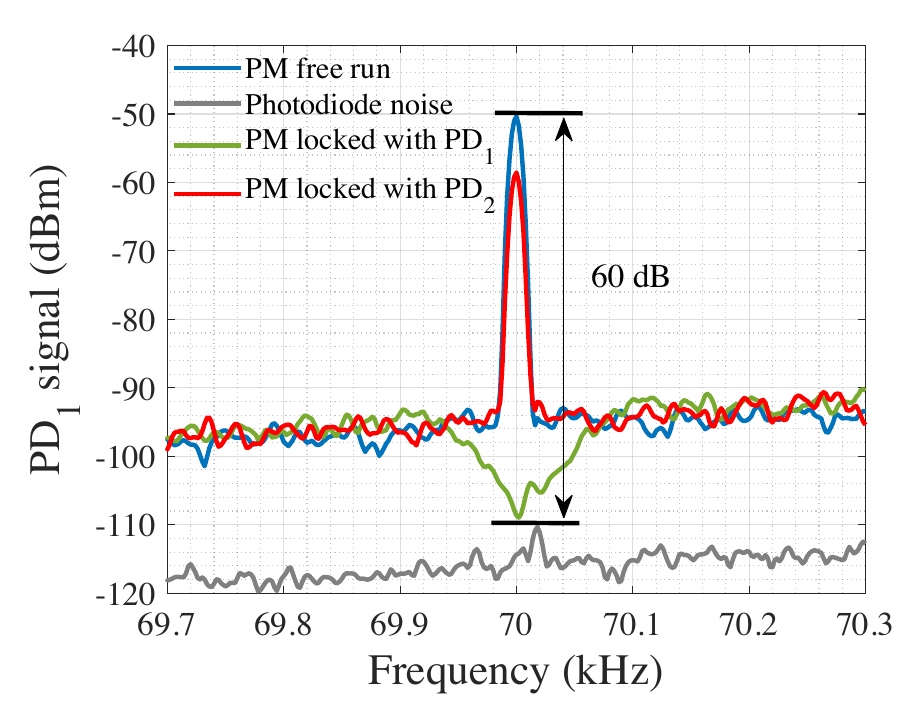}
\caption{Spectra of the photodetector signal of PD$_1$ when the laser frequency is locked using the signal from PD$_1$. Blue line: no RAM control. Red line: RAM reduction using the signal from PD$_2$. Green line:  RAM reduction using the signal from PD$_1$. \textcolor{black}{Gray line: photodiode detection noise.}}
\label{figure-9}
\end{figure}

Figure\,\ref{figure-9} reproduces the spectrum of the signal detected by PD$_1$ while the laser frequency is locked. The blue curve corresponds to the situation where the RAM is not controlled (the RAM servo-loop is open). The large peak at 70~kHz evidences the strong amplitude of this RAM. The red spectrum corresponds to the situation where the RAM is canceled by closing the servo-loop with the signal provided by PD$_2$. The RAM reduction is pretty poor (less than 10~dB), showing that canceling the RAM on PD$_2$ does not at all ensures that the RAM is canceled on PD$_1$. On the contrary, when the demodulated signal $V_{\varepsilon}(\varphi _{\mathrm{d}}+\pi/2)$ is based on the voltage provided by PD$_1$, the green curve shows that the RAM cancellation is extremely efficient (about 60~dB). This shows that, in this situation, the same photodiode signal can be used to lock both the frequency detuning and the RAM amplitude to zero, thus ensuring that the frequency offset due to the RAM vanishes, as evidenced in the following subsection.

\subsection{Measurement of frequency drifts with and without RAM}

\begin{figure}[h!]
\centering\includegraphics[width=11cm]{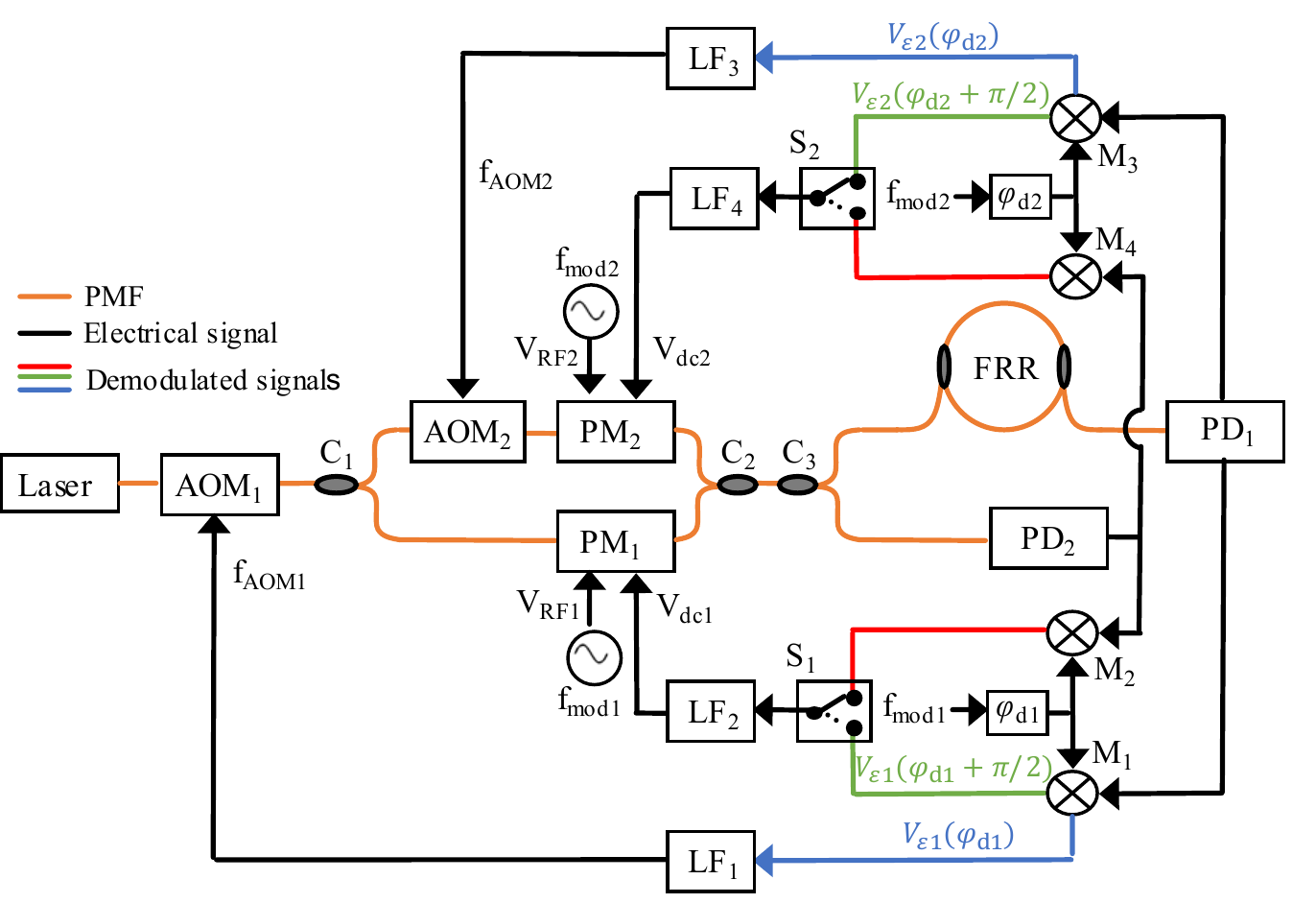}
\caption{Schematic diagram of the FSR measurement of the FRR cavity with the classical and new method of RAM compensation. C: coupler, PD: photodetector, M: mixer, S: switch, LF: loop filter.}
\label{figure-10}
\end{figure}

In order to prove that the new RAM cancellation method permits to get rid of the frequency offsets due to the RAM, we perform the experiment schematized in Fig.\,\ref{figure-10}. In this experiment, we derive two different frequencies from the same laser, and lock them to two different longitudinal modes of the ring resonator \textcolor{black}{in the same direction}. We then use the evolution of the difference between these two optical frequencies to assess the stability of our frequency stabilization loops.

The resonator used in Fig.\,\ref{figure-10} is the same as in the preceding sections. It is probed in transmission in the same propagation direction by two optical frequencies $\nu_{\mathrm{L}}+f_{\mathrm{AOM_1}}$ and $\nu_{\mathrm{L}}+f_{\mathrm{AOM_1}}+f_{\mathrm{AOM_2}}$, where $\nu_{\mathrm{L}}$ is the laser frequency and $f_{\mathrm{AOM_1}}$ and $f_{\mathrm{AOM_2}}$ the frequencies driving the acousto-optic modulators (AOM). The two optical frequencies are locked on two cavity resonances separated by 8 times the cavity FSR, i. e., approximately 108.8~MHz. To this aim, we use two frequency locking servo-loops based on two phase modulators PM$_1$ and PM$_2$ modulated at $f_{\mathrm{mod1}}=70\,\mathrm{kHz}$ and $f_{\mathrm{mod2}}=80\,\mathrm{kHz}$, respectively. The component at $f_{\mathrm{mod1}}$ of the signal produced by PD$_1$ is used to lock $\nu_{\mathrm{L}}+f_{\mathrm{AOM_1}}$ at resonance by controlling $f_{\mathrm{AOM_1}}$. Similarly, the component at $f_{\mathrm{mod2}}$ of the signal produced by PD$_1$ is used to lock $\nu_{\mathrm{L}}+f_{\mathrm{AOM_1}}+f_{\mathrm{AOM_2}}$ at resonance by controlling $f_{\mathrm{AOM_2}}$. For both optical frequencies, the demodulation phases are adjusted to maximize the slopes of the frequency error signals. The control loops LF$_1$ and LF$_3$ here contain digital voltage-to-frequency converters generated by a digital lock-in amplifier from Zurich Instruments (see the details in the Supplemental).

Each of these  frequency locking servo-loops is accompanied by its RAM reduction loop. For both frequencies, we compare two different RAM reduction loops: the classical one based on a detector different from the one used to monitor the frequency, namely PD$_2$ in Fig.\,\ref{figure-10}, or the same one as the one used to control the frequency, namely PD$_1$. In this latter case, the RAM control signals are obtained by demodulating the signals in quadrature with respect to the frequency control signals.
\begin{figure}[h!]
\centering\includegraphics[width=11cm]{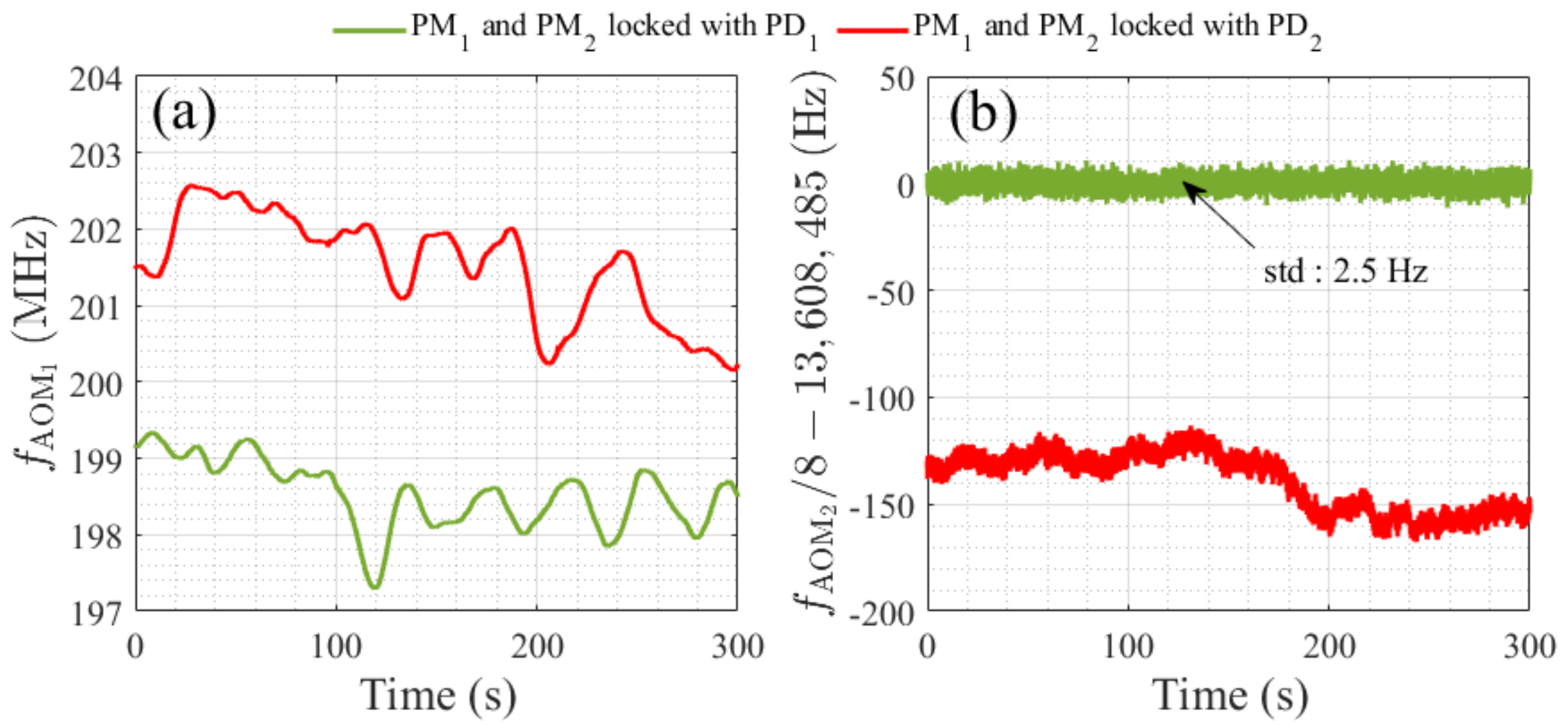}
\caption{(a) Evolution of driven frequency $f_{\mathrm{AOM_1}}$ (b) FSR measurement corresponding to $f_{\mathrm{AOM_2}}/8$ around its mean value 13,608,485 Hz. The red and green curves correspond respectively to the classical and new RAM cancellation methods.}
\label{figure-11}
\end{figure}

The experimental results are shown in Fig.\,\ref{figure-11}. Figure\,\ref{figure-11}(a) compares the evolution over 5 minutes of $f_{\mathrm{AOM}_1}$ when PD$_1$ (green line) and PD$_2$ (red line) are used to cancel the RAMs of the two signals. No clear difference between the two results can be seen, as expected by the fact that in both cases the signal evolves because of the relative drift of the laser frequency with respect to the cavity resonance frequency. This is in strong contrast with the plot of Fig.\ref{figure-11}(b), which reproduces the evolution of $f_{\mathrm{AOM}_2}/8$ for the two RAM cancellation strategies. Once locked, the frequency of AOM$_2$ should be strictly equal to $8\,\Delta\nu_{\mathrm{FSR}}$. The fluctuations in the red curve of Fig.\ref{figure-11}(b) show that the frequency offsets due to the RAMs of the two signals are not perfectly cancelled and fluctuate with time. On the contrary, the green signal is very stable because it is obtained with perfectly cancelled RAMs, and is immune from such fluctuations. Actually, this signal constitutes a very good measurement of the cavity FSR, whose stability is ensured by the thermal stabilization of the resonator. On the contrary, the fluctuations of the red signal cannot be explained by the fluctuations of the optical length of the resonator but only by the fluctuations of the difference between the offsets induced by the RAMs at $f_{\mathrm{mod}1}$ and $f_{\mathrm{mod}2}$.

The fact that the green curve of Fig.\,\ref{figure-11}(b) constitutes a very accurate measurement of the resonator FSR is also confirmed by the value of the standard deviation of this measurement (2.5~Hz in this case). Moreover, we performed the same measurement for many different values of the demodulation phases used to extract the error signals, and obtained a peak-to-peak variation of the measured value of $\Delta\nu_{\mathrm{FSR}}$ of 1~Hz around 13,608,485~Hz, i.e., a relative variation smaller than $10^{-7}$, which compares well with already published results \cite{Mandridis:10,Wang:18}.


\section{Conclusion}

In conclusion, we have proposed and experimentally validated a new RAM cancellation technique. It works in the case of fibered optical systems, for which the classical method introduced in 1985 by Wong and Hall fails due to the  polarization dependent losses \textcolor{black}{and the unavoidably limited extinction ratio of the polarizer located after the PM}. To circumvent this problem, we have shown that the photodiode signal used to provide the frequency error signal necessary to lock the laser at resonance of the considered cavity can also provide an error signal to cancel the RAM. The two error signals are obtained by demodulating with two different phases, for example in quadrature. This has shown to permit to monitor and cancel the RAM at the very place where the signal aiming at locking the laser frequency is produced. 

This new technique has been experimentally validated in a fibered resonator experiment\textcolor{black}{, in the even worse case where the phase modulator is not followed by an output polarizer}. In particular, it has allowed to lock two different frequencies derived from the same laser on two different longitudinal modes of the same cavity, leading to a measurement of the cavity free spectral range with a relative reproducibility and stability of the order of $10^{-7}$ over five minutes. This constitutes an improvement of a factor of about 50 with respect to the standard RAM cancellation technique.

\textcolor{black}{Furthermore, in the present paper, we have limited our discussion to the case of the RAM created by spurious polarization effects. However many other factors can contribute to the RAM. One may then wonder whether the present technique could also be used to provide error signals suited to these situations.}

Such a technique, applicable to resonators probed either in transmission or in reflection, opens interesting perspectives for applications to sensors based on fibered resonators such as, e. g., resonant fiber optic gyroscopes, in which the measurement of minute resonance frequency variations immune from variations of the RAM is of paramount importance.

\section*{Acknowledgments}
This work was performed in the framework of the joint lab between LuMIn and Thales.  We thank Matthieu Goulard, Bertrand Morbieu and Frédéric Seguineau  (Thales AVS) for their help with the fiber cavity. We also thank Alexia Ravaille (Thales Underwater System), Ouali Acef (Observatoire de Paris), Frédéric Du-Burck (Université Sorbonne Paris Nord), and Romain Stomp (Zurich Instruments) for  fruitful discussions, and Erell Laot for checking the calculations.
	
%

\end{document}